\documentclass[final,1p,times]{elsarticleTC}

\usepackage{amsmath,amssymb,amsfonts,amscd}
\usepackage{graphicx}

\usepackage{color,soul}

\begin{document}

\begin{frontmatter}

\title{Graphical representation of marginal and underlying probabilities in quantum mechanics}
\author
{Taksu Cheon}
\ead{taksu.cheon@kochi-tech.ac.jp}
\address
{Laboratory of Physics, Kochi University of Technology,
Tosa Yamada, Kochi 782-8502, Japan}
%

\date{\today}

\begin{abstract}
Wigner's marginal probability theory is revisited, and systematically applied to n-particle correlation measurements.  A set of Bell inequalities whose corollaries are Hardy contradiction and its generalisation are derived with intuitive graphical analysis.
\end{abstract}


\end{frontmatter}

\section{Introduction}
An unambiguous signal of the departure from local causality in quantum mechanics is expressed as violation of Bell inequality \cite{BE65} and its variants, the Bell theorems, such as CHSH inequalities \cite{CHSH}, GHZ theorem \cite{GH90} and Hardy\rq s arguments \cite{HA92, HA93}, which has been verified by the experiments by use of various physical systems \cite{Aspect82, Weihs98, Rowe01, Sakai06, Go07}.
The relations among these Bell theorems have been uncovered gradually \cite{Cereceda01, Cereceda04}.
The violation of Bell inequality can be viewed also as a sign of the breakdown of the classical concepts of joint probability.
As independently pointed out by Wigner and Fine \cite{WI70, FI82a, FI82b}, the Bell inequality, in its original form devised by Bell, presupposes underlying joint probability distribution for all possible measurement setups and outcomes in the experiment, whose marginal probabilities yield the probability distributions of the actual measurement setup of each run. 

In this article, we revisit the argument by Wigner and Fine \cite{WI70, FI82a, FI82b}, and try to make clear the relation between their arguments and the concept of local reality.  In the process, we uncover the graphic structure of the underlying joint probability, which we utilise to systematically to search general form of Hardy's equalities and associated inequalities.

This article is organized as follows. In Sec.~2, we introduce the concepts of the underlying probabilities which respects statistical independence and yield the measurable probability distributions as their marginal probabilities. 
In Sec.~3, we check consistency between the assumptions of statistical independence and the local reality. 
In Sec.~4, we turn to the graphical representation of the marginal probability for two-by-two experiment. Using this representation, we clarify that Hardy\rq s equalities can be thought of as an extremal case of CHSH inequality. In Sec.~5, we generalise the above observation for multi-setting case and multi-partite case. Sec.~5 has the discussion and our conclusion.

\section{Marginal and Underlying Probabilities, Statistical Independence of Measurements}
Consider a system of two 1/2-spins,
each of which are measured separately by two observers Alice and Bob,
both of whom  can set up the measurement devices along two different axes,
$a = 0, 1$ for Alice , and $b=0, 1$ for Bob.
We assume that the result of the dichotomic measurements is specified by 
$A = 0, 1$ for Alice, and $B = 0, 1$ for Bob.  The
usual sign notations of projection are recovered 
by identifications $s_A = (-)^A$ and $s_B = (-)^B$. 

Let us denote the probability of Alice finding the system in
the state $A$ along the $a$-axis and Bob in $B$ 
along $b$-axis by $P_{ab}(A, B)$.
The assumption of statistical independence is expressed as
\begin{eqnarray}
\label{stindep1}
P_{ab}(A, B) = P_{a}(A) P_{b}(B) .
\end{eqnarray}
Let us now consider parallel measurements on an ensemble of the system 
in two combinations of axes pair $(a,b)$ and $(a',b')$.   
We write the joint probability of finding the
results $(A, B)$ and  $(A', B')$ as 
$W([A, B]_{ab}; [A', B']_{a'b'})$.  If the parallel measurements are
statistically independent, we should have
\begin{eqnarray}
\label{e2}
W([A,B]_{ab}; [A',B']_{a'b'}) 
= P_{ab}(A,B) P_{a'b'}(A',B') ,
\end{eqnarray}
From the assumption (\ref{stindep1}),  we have
\begin{eqnarray}
\label{e3}
P_{00}(A_0,B_0) P_{11}(A_1,B_1) 
= P_{01}(A_0, B_1) P_{10}(A_1, B_0)
\end{eqnarray}
since both are given by the product 
$ P_{0}(A_0) P_{1}(A_1) P_{0}(B_0) P_{1}(B_1)$.  
This immediately leads to the equivalence 
between $W([A_0, B_0]_{00}; [A_1, B_1]_{11})$ 
and $W([A_0, B_1]_{01}; [A_1, B_0]_{10})$.
This means that we can define 
unconditional underlying probability $\rho(A_0, B_0; A_1, B_1)$
which is the joint probability of 
Alice finding her system in state $A_0$ along $0$-axis,
$A_1$ along $1$-axis and 
Bob finding his particle in state $B_0$ along $0$-axis,
$B_1$ along $1$-axis;
\begin{eqnarray}
\label{udef}
\rho(A_0, A_1; B_0, B_1) 
=W([A_0, B_0]_{00}; [A_1, B_1]_{11})
=W([A_0, B_1]_{01}; [A_1, B_0]_{10}) ,
\end{eqnarray}
which signify the ``reality'' of the set of physical observables
$(A_0, B_0 ; A_1, B_1)$ irrespective to the measurement.
There are $2^4=16$ of $U$s.   As probabilities, they are all non-negative real numbers,
which add up to unity.
The observable probabilities $P_{ij}$ are obtained from $\rho$ by partial sum as
\begin{eqnarray}
\label{u2p}
&&
P_{00}(A_0,B_0)=\sum_{A_1,B_1}\rho(A_0, B_0; A_1, B_1) , 
\quad
P_{10}(A_1,B_0)=\sum_{A_0,B_1}\rho(A_0, B_0; A_1, B_1) , 
\nonumber \\ &&
P_{01}(A_0,B_1)=\sum_{A_1,B_0}\rho(A_0, B_0; A_1, B_1) , 
\quad
P_{11}(A_1,B_1)=\sum_{A_0,B_0}\rho(A_0, B_0; A_1, B_1).
\end{eqnarray}
Following Wigner, we call the probability $P_{ab}$s that are obtainable from 
direct measurements as {\it marginal probabilities}.
Although the existence of unconditional probability $\rho$ which is 
guaranteed by the
relations (\ref{udef}), or equivalently (\ref{e3}), are derivable from 
the statistical independence between Alice's and Bob's observables, (\ref{stindep1}), 
the former is a looser assumption, from which the latter may not necessarily follow. 
So irrespective to the logic we have followed until now,
we shall use the existence of unconditional probability $U$ expressed in relations
(\ref{udef}) and (\ref{u2p}) as the basic assumption.
\section{Derivation of Joint Probability Based on Locally Realistic Theory with Hidden Variable }

Although in the derivation of
$\rho(A_0, B_0; A_1, B_1)$, the cornerstone of our approach, we have pretended
that it is derived from the statistical independence assumption of marginal probilities, 
it can be derived by quite deferent set of assumption based on the concept of
{\it hidden variable theory with local realism}.
Of course, this is how the Bell inequality has been historically derived, and 
this is exactly the reason that Bell theorem is termed one of the most profound theorem
in phyisics.
We need, however to put it in perspective in view of the fact that there are vocal dissenting
view \cite{FI96} on the significance of Bell experiment that what is proven 
is just the statistical separability of probabilities, not the negation of local realism. 

Here, we detail how Wigner has arrived at conditional and unconditional
underlying probabilities $W$ and $\rho$, stating from deterministic theory 
with hidden variables.
Although we base our argument on the specific case of two-by-two Bell 
experiment, readers will see its generality and applicability 
to broader situations.
We first want to construct a framework of deterministic theory
with hidden variables, that are,
unknown variables which are not to be observed directly, but
whose ensemble average generates all $2^4=16$ marginal
probabilities $P_{ab}(A,B)$, that completely specifies the outcomes of
two-by-two Bell experiment.  

It is, in fact, rather easy to have such a theory,
given sufficient number of variables are brought in.
Consider four variables $q_{ab}$ with $a = 0, 1$ and $b = 0, 1$.
we assume $q_{ab}$ takes four values 0,1,2 and 3, which we express as
\begin{eqnarray}
q_{ab} = A_{ab}+2B_{ab}
\end{eqnarray}
with binary variables $A_{ab}=0,1$ and $B_{ab}=0,1$.  We can alternatively think
a {\it byte} variable  $\lambda$ made up of eight bits $ \{ A_{ab}, B_{ab}\}$, namely
\begin{eqnarray}
\Lambda = \{A_{00},B_{00},A_{10},B_{10},A_{01},B_{01},A_{11},B_{11} \}
\end{eqnarray}
as our hidden variables.  These variables are
governed by some unspecified dynamics.  It is just sufficient if we accept that
the result of the projection measurement with axes $a$ for Alice and $b$
for Bob is determined by the value of
$q_{ab}$ in such way that the projections $s_A$ and $s_B$ for Alice
and Bob respectively are given by $s_A = (-)^{A_{ab}}$
and $s_B = (-)^{B_{ab}}$.

We now consider an ensemble of this deterministic system whose density of distribution on the space of $\Lambda$ is given by 
\begin{eqnarray}
W(\Lambda) =W( A_{00},B_{00};A_{10},B_{10};A_{01},B_{01};A_{11},B_{11} )
\end{eqnarray}
The observable probability $P_{ab}$ are given by
\begin{eqnarray}
&&\!\!\!\!\!\!\!\!\!\!
P_{00}(A,B) = 
\sum_{ A_{10},B_{10}}\sum_{A_{01},B_{01}}\sum_{A_{11},B_{11}} 
 {W(  A,B;A_{10},B_{10};A_{01},B_{01};A_{11},B_{11}  )} ,
\nonumber \\
&&\!\!\!\!\!\!\!\!\!\!
P_{10}(A,B) = 
\sum_{ A_{00},B_{00}}\sum_{A_{01},B_{01}}\sum_{A_{11},B_{11}} 
 {W(  A_{00},B_{00}; A,B; A_{01},B_{01};A_{11},B_{11}  )} ,
\qquad etc.,
\end{eqnarray}
and we should always be able to construct a deterministic theory 
whose initial ensemble of variable $\Lambda(t=0)$,
that corresponds to the initial setup of the experiment,
evolve into the distribution $W(\Lambda)$ with 
$\Lambda=\Lambda(t)$ at the time of the measurement, 
that reproduces the observed $P_{ab}(A,B) $. 
It has to be remarked that
the existence of this supposed underlying theory itself is rather unremarkable.
The fact, that it requires an eight-bits variables capable of $2^8=256$
values to reproduce $16$ experimental numbers, makes this theory
an example of ``cost-ineffective'' generalization.

Now, we consider only those deterministic theories that
respect local realism {\it {\` a} la}  J.S.Bell.  Then, the result of the Alice's
measurement should not depend on {\it how Bob's measurement
device is placed}, and Bob's result should not depend on 
{\it how Alice's measurement device is placed} .
The values of the variables, in this type of theory, have to be limited
in such way to reflect this fact, namely
\begin{eqnarray}
A_{00}=A_{01} \equiv A_0,
\quad
A_{10}=A_{11} \equiv A_1,
\quad
B_{00}=B_{10} \equiv B_0,
\quad
B_{01}=B_{11} \equiv B_1 .
\end{eqnarray}
So half of the bits in the variable $\Lambda$ are redundant. 
In other word, local realistic theories underlying the observable
$P_{ab}(A,B)$ is to be described by a {\it half byte} variable $\lambda$,
which we define in the form
\begin{eqnarray}
\lambda = \{A_{0},B_{0},A_{1},B_{1} \} .
\end{eqnarray}
The ensemble-averaged system we observe in experiments
is to be specified by the density of distribution
\begin{eqnarray}
\rho(\lambda) =\rho( A_{0},B_{0}; A_{1},B_{1} ) ,
\end{eqnarray}
which gives the observable probability $P_{ab}$ in the form
\begin{eqnarray}
P_{00}(A,B) = \sum_{ A_{1},B_{1}}  { \rho(  A, B; A_{1}, B_{1}) } ,
\quad
P_{10}(A,B) = \sum_{ A_{0},B_{1}}  { \rho(  A_{0}, B; A, B_{1}) } ,
\qquad etc..
\end{eqnarray}

Note that the variable $\lambda$ can take 16 discrete values,
as opposed to 256 for $\Lambda$.  Consequently, there are 16 of 
$\rho(\lambda)$ functions as in contrast to 256 of $W(\Lambda)$ functions.

In formal term, the local realistic subclass of hidden variable theory is
obtained from all possible theory by the reduction of density distribution
\begin{eqnarray}
W( A_{00},B_{00};A_{10},B_{10};A_{01},B_{01};A_{11},B_{11} )
 =\rho( A_{00},B_{10}; A_{11},B_{01} )
 \delta_{A_{00},A_{01}} \delta_{A_{01},A_{11}}
 \delta_{B_{00},B_{10}} \delta_{B_{01},B_{11}},
\end{eqnarray}
or conversely,
\begin{eqnarray}
\rho( A_{0}, B_{0}; A_{1}, B_{1} )
=W( A_{0}, B_{0}; A_{1},B_{0}; A_{0},B_{1}; A_{1},B_{1} ) .
\end{eqnarray}

We see that the \lq\lq hidden variables\rq\rq~ $A_{ab}$ , $B_{ab}$ and their reductions 
$A_{a}$ , $B_b$ are indeed hidden behind 
their guise as projection indices in $W$ and $U$. 

\section{Diagrammatical Proof of Bell inequalities and Hardy Contradiction}
The assumption of the existence of 
underlying unconditional probability $\rho(A_0, B_0; A_1, B_1)$
leads to several relations among marginal probabilities $P_{ab}(A,B)$.

Let us first reconfirm our notational conventions.  
The index $a$, that takes two values 0 and 1, stands for the axis of choice by Alice,
along which the first particle is projectively measured to yield the value $s_A=+$ or
$-$, which is alternatively expressed as $A = 0$ or $1$.
Similarly, the index $b$, that can be 0 or 1, stands for the axis of Bob's measurement,
which yields the value $s_B = +$ or $-$ that is alternatively expressed as
$B = 0$ or $1$.
%
%
We define inter-particle axis specific projections 
\begin{eqnarray}
\label{alphadef}
& &\nonumber \\
i = A_0 + 2 B_0 , 
\qquad
j = A_1 + 2 B_1 .
\end{eqnarray}
The value of  $i$ represents the combined projection 
along the ``0'' axes of Alice and Bob,
and $j$ along their ``1''- axes.
We now reorder $\rho(A_0, A_1; B_0, B_1)$ by the indices $i$ and $j$
with the use of a matrix $V$ defined by
\begin{eqnarray}
\label{vdef}
V_{i,j} 
= \rho(A_0,B_0; A_1, B_1)|_{(i = A_0 + 2 B_0,j = A_1 + 2 B_1)} .
\end{eqnarray}
The partial sum (\ref{u2p}) then becomes 
\begin{eqnarray}
\label{v2p}
&&
P_{11}(A,B) = \sum_{j=0}^3{ V_{A+2B,j} }
\quad
P_{10}(A,B) = \sum_{i=0}^1 \left({ V_{2B+i,A} } +{ V_{2B+i,A+2} } \right), 
\nonumber \\ &&
P_{01}(A,B)= \sum_{j=0}^1\left( { V_{A,2B+j} } + { V_{A+2,2B+j} } \right) , 
\quad
P_{11}(A,B) = \sum_{i=0}^3{ V_{i,A+2B} }
 \end{eqnarray}
These expression has a very intuitive graphical representation Fig. 1.  
If we place $\rho$ on a 4-by-4
grid according to the indices $i$ and $j$,  $P_{00}$ are given by sums 
along horizontal lines,
$P_{11}$ by sums along vertical lines, while $P_{10}$ and $P_{01}$ are given by sums 
ribbon-shaped lines vertically and horizontally placed.

%
\begin{figure}[h]
\center{
\includegraphics[width=12cm]{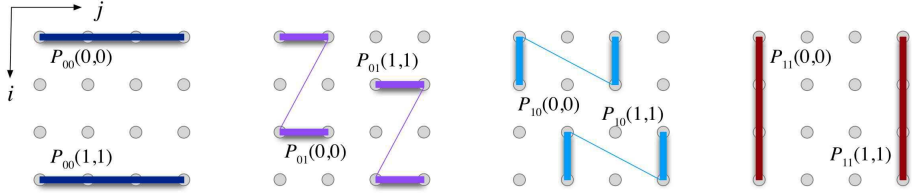}
}
\label{fig1}
\caption
{
Marginal probabilities  $P_{ab}(A,B)$ represented as a line or linked lines which cover the
the grids that represent the underlying probabilities $V_{i,j}$ to be summed up.
}
\end{figure}
From the compositions (\ref{v2p}), it follows, for example, that 
\begin{eqnarray}
\label{hardcmp1}
P_{10}(0,0) + P_{01}(0,0) + P_{11}(1,1)
= &&\!\!\!\!\!\!\!\!
\ 2 V_{0,0} + V_{0,1} + V_{0,2} + V_{0,3}
\nonumber \\ &&\!\!\!\!\!\!\!\!
+ V_{1,0} \qquad\quad\! + V_{1,2} + V_{1,3}
\nonumber \\ &&\!\!\!\!\!\!\!\!
+ V_{2,0} + V_{2,1} \qquad\quad\! + V_{2,3}
\\ \nonumber &&\!\!\!\!\!\!\!\!
\qquad\quad\!\qquad\quad\!\qquad\quad\! \!+\, V_{3,3}
\end{eqnarray}
which contains 
\begin{eqnarray}
\label{hardcmp2}
P_{00}(0,0) = V_{0,0} + V_{0,1} + V_{0,2} + V_{0,3}
\end{eqnarray}
and some more positive quantities.  
Thus we have a Bell inequality in the form
\begin{eqnarray}
\label{hardy00}
P_{10}(0,0) + P_{01}(0,0) + P_{11}(1,1) - P_{00}(0,0) \ge 0
\end{eqnarray}
A graphical representation of this inequality is shown in Fig. 2.

%
%
\begin{figure}
\center{
\includegraphics[width=2.5cm]{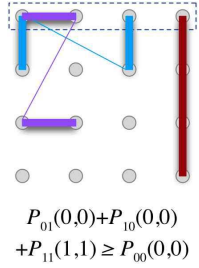}
}
\label{fig2}
\caption
{
Graphical depiction of a Bell inequality among marginal probabilities $P_{ab}(A,B)$ represented as a line or linked lines for terms in LHS, a dashed box for the RHS, superimposed on the grid representing the underlying probabilities $V_{i,j}$ to be summed up.
}
\end{figure}
A corollary immediately obtained is that
\begin{eqnarray}
\label{hardycc}
{\rm if \ \ }
P_{10}(0,0) = P_{01}(0,0) = P_{11}(1,1) = 0 
{\rm \ \ hold} ,
{\rm \ \ then \ \ }
P_{00}(0,0) = 0 
{\rm \ \ follows}.
\end{eqnarray}
This is nothing but the equality obtained by Lucien Hardy.  
A similar construction to (\ref{hardcmp1}), (\ref{hardcmp2}) leads to
\begin{eqnarray}
\label{hardy11}
&&
P_{10}(1,1) + P_{01}(1,1) + P_{11}(0,0) - P_{00}(1,1) \ge 0 ,
\nonumber \\ &&
P_{10}(1,0) + P_{01}(1,0) + P_{11}(0,1) - P_{00}(1,0) \ge 0 ,
\\ \nonumber &&
P_{10}(0,1) + P_{01}(0,1) + P_{11}(1,0) - P_{00}(0,1) \ge 0 .
\end{eqnarray}
With the definition of corelation $C_{ab}$,
\begin{eqnarray}
\label{corrdef}
C_{ab} = P_{ab}(0,0)-P_{ab}(1,0)-P_{ab}(0,1)+P_{ab}(1,1)
\end{eqnarray}
inequalities (\ref{hardy00}) and (\ref{hardy11}) lead to
\begin{eqnarray}
\label{chsh}
&&
|C_{10} + C_{01} + C_{11} - C_{00}| \le 2 ,
\end{eqnarray}
which is a celebrated CHSH inequality.

With the interchanges of axis indices $a$ and $b$, and with interchanges of projections 
$0$ and $1$ for both $A$ and $B$ in $P_{ab}(A,B)$, 
there are 64 inequalities of the type (\ref{hardy00}).  Each four of them 
sharing the same leg ${ab}$ form a single CHSH inequality. 
As an obvious corollary, there are 64 variants of Hardy equality.

\section{Three Particle Generalization and More}
The extension to $n$-particle case is rather straightforward.
We illustrate it with three particle case.
Let us now consider a system consisting of three spin 1/2 particles that
are respectively measured by Alice, Bob and Chris, all of whom have two choices 
each for the orientation of measurement represented by $a=0$ or $1$, $b=0$ or $1$ 
and $c=0$ or $1$, respectively.  
Alice's measured projection is represented by $A_a=0$ or $1$, Bob's by
$B_b=0$ or $1$ and Chris' by $C_c=0$ or $1$.
If there is an underlying probability $\rho(A_0, B_0, C_0; A_1, B_1, C_1)$, 
that presuppose the simultaneous existence of all projections for all three observers, 
the marginal probabilities $P_{abc}$s which are
the joint probabilities of direct observables are given by the partial sums
\begin{eqnarray}
\label{u2p3}
&&
\!\!\!\!\!\!\!\!\!\!\!\!\!\!\!\!
P_{000}(A,B,C)=\sum_{A',B',C'} \rho(A, B, C; A', B', C')  , 
\ \ 
P_{100}(A,B,C)=\sum_{A',B',C' }\rho(A, B, C; A', B', C'),  
\nonumber \\ &&
\!\!\!\!\!\!\!\!\!\!\!\!\!\!\!\!
P_{010}(A,B,C)=\sum_{A',B',C'} \rho(A, B, C; A', B', C'),  
\ \ 
P_{110}(A,B,C)=\sum_{A',B',C'} \rho(A, B, C; A', B', C'),  
\nonumber \\ &&
\!\!\!\!\!\!\!\!\!\!\!\!\!\!\!\!
P_{001}(A,B,C)=\sum_{A',B',C'} \rho(A, B, C; A', B', C'),  
\ \ 
P_{101}(A,B,C)=\sum_{A',B',C'} \rho(A, B, C; A', B', C'),  
\nonumber \\ &&
\!\!\!\!\!\!\!\!\!\!\!\!\!\!\!\!
P_{011}(A,B,C)=\sum_{A',B',C'} \rho(A, B, C; A', B', C'),  
\ \ 
P_{111}(A,B,C)=\sum_{A',B',C'} \rho(A, B, C; A', B', C')  .
\end{eqnarray}
This set of construction may or may not result in the assumption of statistical independence 
\begin{eqnarray}
\label{stindep3}
P_{abc}(A,B,C)
= P_{a}(A) P_{b}(B) P_{c}(C) ,
\end{eqnarray} 
although the latter necessarily results in the former.  In case (\ref{stindep3}) holds,
we have
\begin{eqnarray}
\label{e33}
&&\!\!\!\!\!\!\!
P_{000}(A,B,C) P_{111}(A',B',C') 
= P_{001}(A,B,C') P_{110}(A',B',C) ,
\nonumber \\
&&\!\!\!\!\!\!\!
P_{000}(A,B,C) P_{110}(A',B',C') 
= P_{100}(A',B,C) P_{010}(A,B',C') ,
\nonumber \\
&&\!\!\!\!\!\!\!
P_{111}(A,B,C) P_{001}(A',B',C') 
= P_{011}(A',B,C) P_{101}(A,B',C') ,
\end{eqnarray}
{\it etc.}.
%
%
\begin{figure}
\center{
\includegraphics[width=15cm]{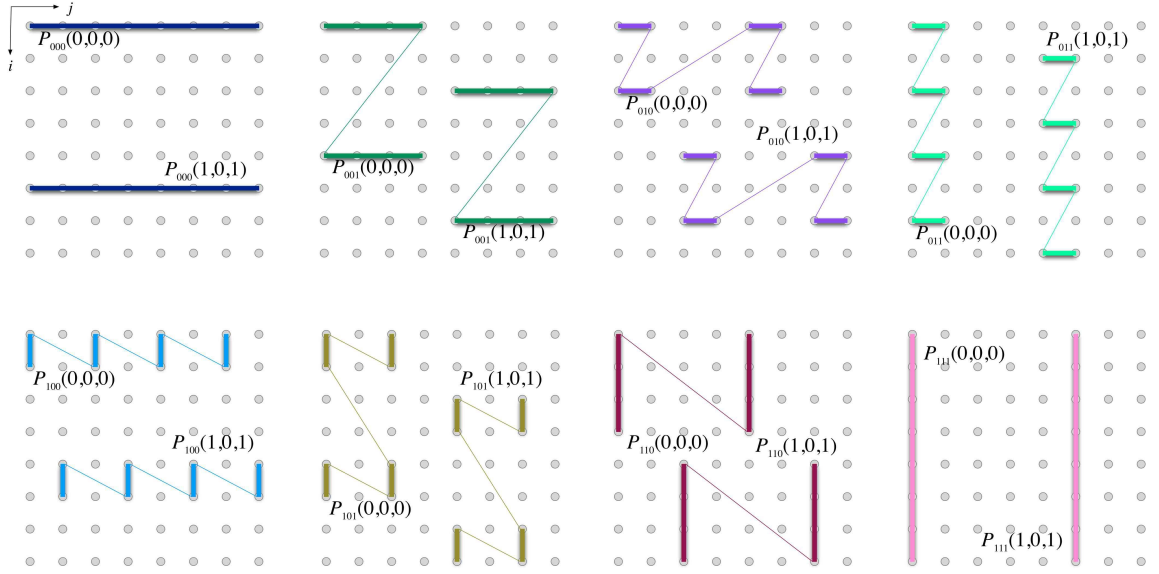}}
\label{fig3}
\caption
{
Graphical depiction of three-body marginal probabilities $P_{ab}(A,B,C)$ represented as a line or linked lines superimposed on the grids representing underlying probability $V_{i,j}$ to be summed up.
}
\end{figure}
With the three-digits grouping of indices
\begin{eqnarray}
\label{vdef3}
V_{i,j} 
= \rho(A_0, B_0, C_0; A_1, B_1, C_1)
  |_{(i = A_0+2B_0+4C_0, j = A_1 + 2B_1+4C_1)} ,
\end{eqnarray}
the partial sums (\ref{u2p3}) become 
\begin{eqnarray}
\label{v2p3}
&&
P_{000}(A,B,C)=\sum_{j=0}^7 { V_{A+2B+4C,j} } , 
%
\quad
P_{111}(A,B,C)=\sum_{i=0}^7 { V_{i,A+2B+4C} } ,  
\nonumber \\ 
&&\!\!\!\!\!\!\!\!\!\!\!\!\!\!
P_{110}(A,B,C)=\! \sum_{i=0}^3 \left( { V_{4C+i,A+2B} }+ \!{ V_{4C+i,A+2B+4} }  \right) ,
%
\ 
P_{001}(A,B,C)=\! \sum_{j=0}^3 \left( { V_{A+2B,4C+j} }+ \!{ V_{A+2B+4,4C+j} }  \right) , 
\nonumber \\
&&\!\!\!\!\!\!\!\!
P_{100}(A,B,C)
=\sum_{i=0}^1 \left( { V_{2B+4C+i,A} }+ { V_{2B+4C+i,A+2} }
+ { V_{2B+4C+i,A+4} }+ { V_{2B+4C+i,A+6} } \right) ,
\nonumber \\ 
&&\!\!\!\!\!\!\!\!
P_{011}(A,B,C)
=\sum_{j=0}^1 \left( { V_{A,2B+4C+j} }+ { V_{A+2,2B+4C+j} }
 +{ V_{A+4,2B+4C+j} }+ { V_{A+6,2B+4C+j} }  \right) ,
 \nonumber \\
&&\!\!\!\!\!\!\!\!
P_{010}(A,B,C)
=\sum_{j=0}^1 \left({ V_{A+4C,2B+j} }+ { V_{A+4C+2,2B+j} }
+{ V_{A+4C,2B+4+j} }+ { V_{A+4C+2,2B+4+j} } \right) ,
\nonumber \\ 
&&\!\!\!\!\!\!\!\!
P_{101}(A,B,C)
=\sum_{i=0}^1 \left( { V_{2B+i,A+4C} }+ { V_{2B+i,A+4C+2} }
+  { V_{2B+4+i,A+4C} }+ { V_{2B+4+i,A+4C+2} }  \right) .
%
\end{eqnarray}
This extends the previous graphical expression of 4-by-4 matrix
grid for two-particle case with 8-by-8 matrix grid on which $\rho$s are placed with
indices $i=A_0+2B_0+4C_0$ and $j= A_1+2B_1+4C_1$.
$P_{000}$ and $P_{111}$ are sums over horizontal and vertical lines, while
$P_{100}$, $P_{010}$ and $P_{001}$ are sums over variously shaped ribbons,
and $P_{011}$, $P_{101}$ and $P_{110}$ their respective mirror images 
with respect to the diagonal lines, all indicated in Fig.3.
Comparison between Fig. 1 and Fig. 3 reveals how the $n=3$ graphs are made 
out of $n=2$ graphs;  $P_{000}$, $P_{100}$ and $P_{010}$ are just 
the "sideway doubling" of $P_{00}$, $P_{10}$ and $P_{01}$, and the
remaining $P_{111}$, $P_{011}$ and $P_{101}$ are obtained by
mirroring with respect  to diagonal line that corresponds to the reversing 
axis indices ($a,b,c$: $0 \leftrightarrow 1$). 
This construction carries over to any $n \to n+1$ extension. 
%
%
Drawing various ribbons on the matrix grid as before, the following inequality 
is shown to hold
\begin{eqnarray}
\label{hardy300}
P_{100}(0,0,0) + P_{010}(0,0,0) + P_{001}(0,0,0) + P_{111}(1,1,1) - P_{000}(0,0,0) \ge 0
\end{eqnarray}
whose corollary is a three body extension of Hardy equality, which states
\begin{eqnarray}
\label{hardy3cc}
{\rm if \ \ }
P_{100}(0,0,0) = P_{010}(0,0,0) = P_{001}(0,0,0) = P_{111}(1,1,1) = 0 
{\rm \ \ hold} , 
\nonumber \\
{\rm \ \ then \ \ }
P_{000}(0,0,0) = 0 
{\rm \ \ follows}.
\end{eqnarray}
%
%
%
\begin{figure}[h]
\center{
\includegraphics[width=4cm]{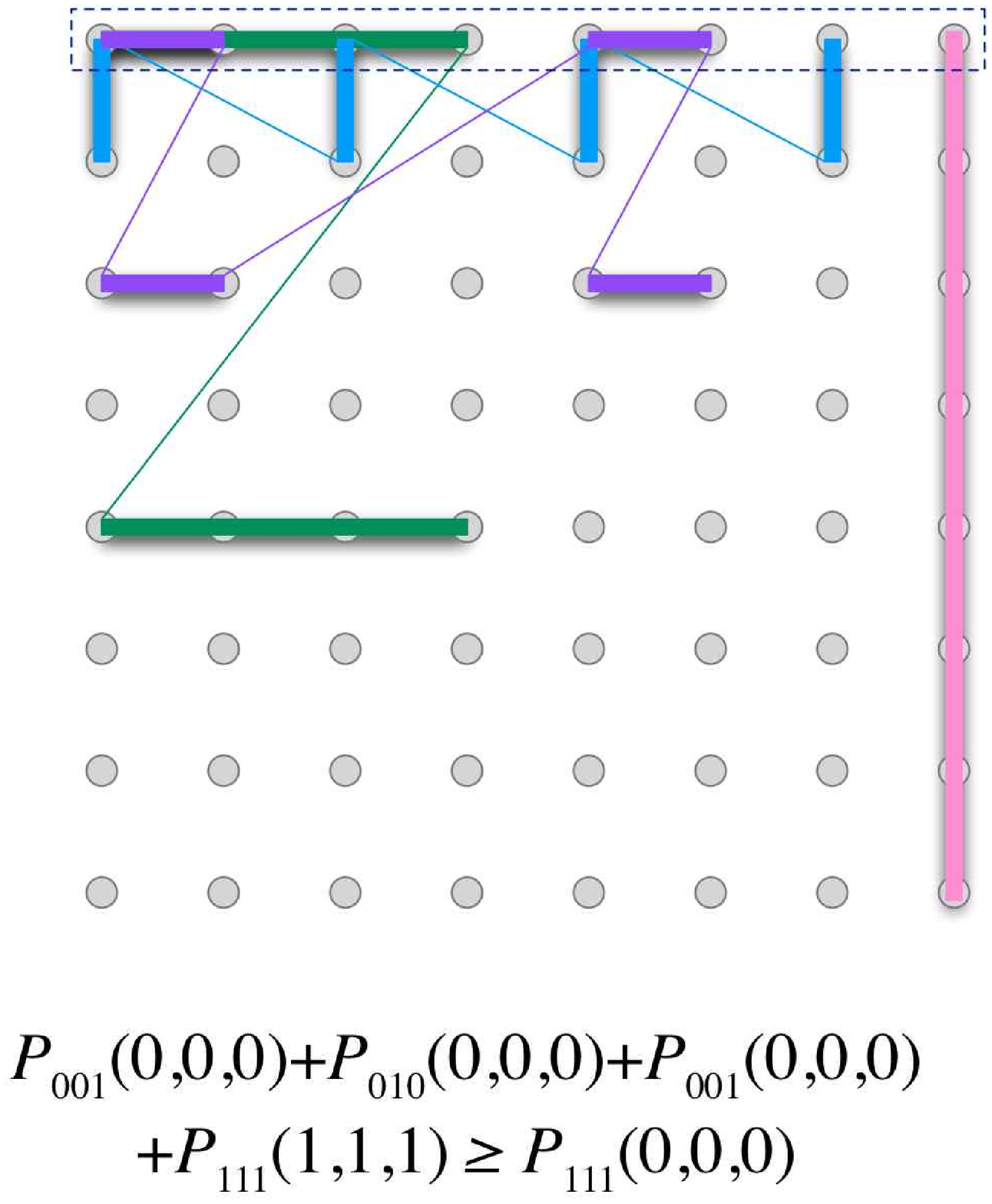}}
\label{fig4}
\caption
{
Graphical depiction of three-body marginal probabilities for the extended Hardy contradiction, represented as a line or linked lines for terms in LHS, a dashed box for the RHS, superimposed on the grid representing the underlying probabilities $V_{i,j}$ to be summed up.
}
\end{figure}
The geralization of this result 
\begin{eqnarray}
\label{hardy-n}
&&\!\!\!\!\!\!\!\!\!\!\!\!\!\!\!\!
P_{100...0}(0,0,...,0) + P_{010...0}(0,0,...,0) +...+ P_{00...01}(0,0,...,0)  
\nonumber \\ &&\qquad\qquad\qquad\qquad
+ P_{11...1}(1,1,...,1) - P_{00...0}(0,0,...,0) \ge 0 ,
\end{eqnarray}
and its associated Hardy equality is not hard to prove.
%
%
%
%
\begin{figure}
\center{
\includegraphics[width=13cm]{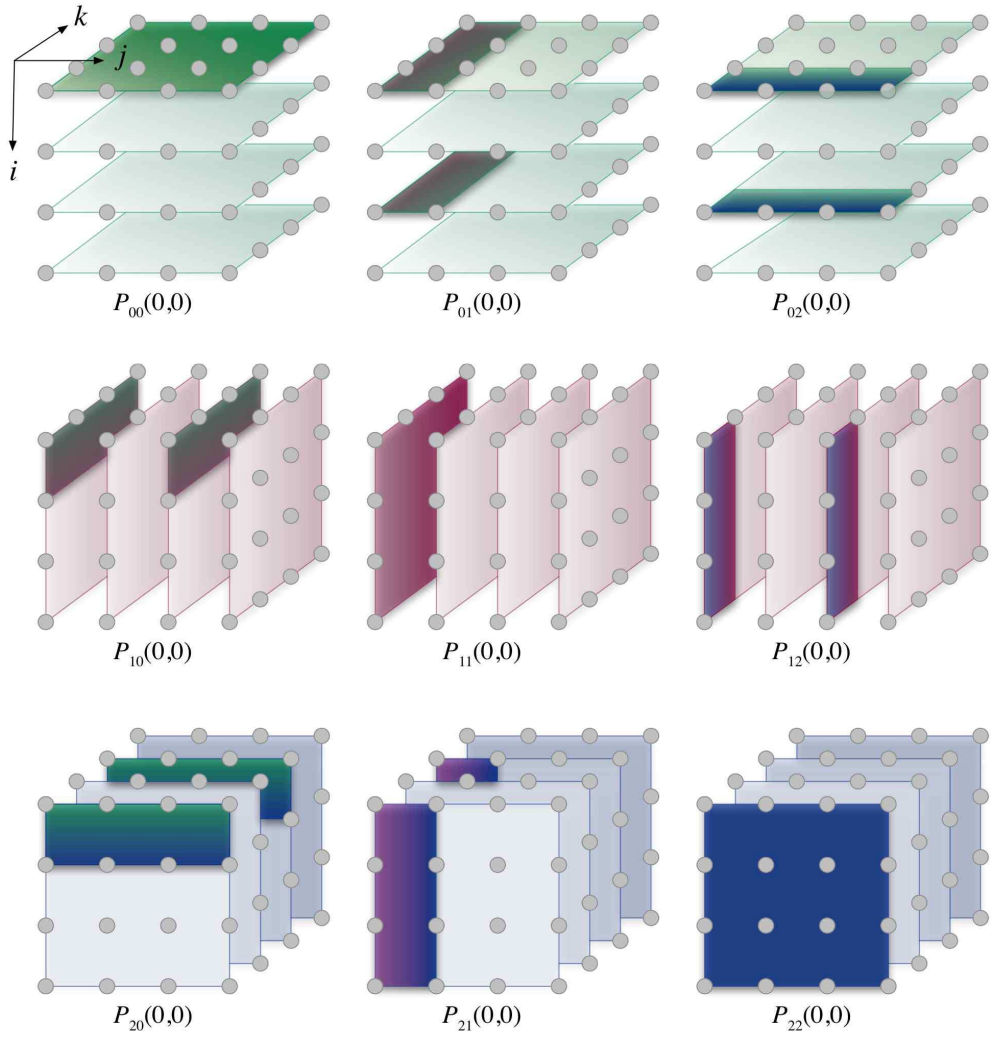}}
\label{fig5}
\caption
{
Graphical depiction of two-body three-axes marginal probabilities $P_{ab}(A,B)$ represented as planes painted in identical colors, which cover the three-dimensional grids that represent the underlying probabilities $V_{i, j, k}$ to be summed up.}
\end{figure}
Quite separately from above extension, we can prove a Zukowski inequality
\begin{eqnarray}
\label{zukow}
C_{111}-C_{001}-C_{010}-C_{100} \ge -2 
\end{eqnarray}
in a similar fashion with the ribbons on 8-by-8 grid, whose corollary
\begin{eqnarray}
\label{GHZ}
{\rm if \ \ }C_{001}=C_{010}=C_{100}=1 {\rm \ \  then \ \ } C_{111}=1 
\end{eqnarray}
is, of course, the negation of GHZ contradiction.
Another set of extension exists in the form of two party $m$-axes Bell experiments.
Take for example, a system of two spin 1/2 particle each measured by Alice and Bob 
both of whom now has a choice of three projection axes $a=0, 1, 2$ and $b=0, 1, 2$.
Experimental results are specified by 64 marginal probabilities $P_{ab}(A,B)$. 
The assumption of underlying unconditional probability distribution now involves 
64 quantities
$\rho(A_0, B_0; A_1, B_1; A_2, B_2)$.
The proper tabulation of $\rho$ now requires the definition of three two-digits indices
\begin{eqnarray}
\label{ijk}
i_k = A_k + 2B_k,
\quad
k=0, 2, 3,
\end{eqnarray}
thus is a cube grid of size 4-by-4-by-4.
As depicted in Fig. 5,
$P_{kk}$ is the sum over a slice parallel to the cube surfaces, and $P_{kl}$ with $k \ne l$
half sums of two non adjacent surfaces.

In a similar fashion to the previous cases, although
now requiring some 3-dimensional recognition of patterns, we can prove a set of
three-axes nonlocality inequalities.  One such example is depicted in Fig. 6, showing 
an inequality
\begin{eqnarray}
\label{hardy33}
P_{00}(1,1) + P_{10}(0,0) + P_{02}(0,0)   - P_{12}(0,0) \ge 0 .
\end{eqnarray}
This is a three-axes version of inequalities of the type (\ref{hardy00}).  In a sense, 
it can be regarded as 
physically identical to them, since just by renaming the axis "2" of Bob 
as "1", this simply reduces back to one of the two-axis type inequality.
But this inequality does involve genuinely different three axes. 
We may obtain a three-axes version of Hardy type equality again by
setting the fisrt three term of  (\ref{hardy33}) to be zero.
Instead, we show another interesting face of this inequality by 
{it requiring only the first term to be zero}.
We then have
\begin{eqnarray}
\label{bellbell}
{\rm if\ \ } P_{00}(1,1) = 0, {\rm \ \ then \ \ }
P_{10}(0,0) + P_{02}(0,0)   - P_{12}(0,0) \ge 0 .
\end{eqnarray}
This is nothing but the original Bell inequality in the form devised by J. S. Bell
expressed in marginal probabilities instead of correlation functions,
and the proof shown here is just the graphical dressing of Wigner's proof.
Note that the original requirement of ``being in singlet state'' is loosened to 
$ P_{00}(1,1) = 0$.  We can now see that ``Bell's'' Bell inequality occupies 
a midpoint between CHSH type inequality and Hardy type equality.

%
%
%
\begin{figure}
\center{
\includegraphics[width=4cm]{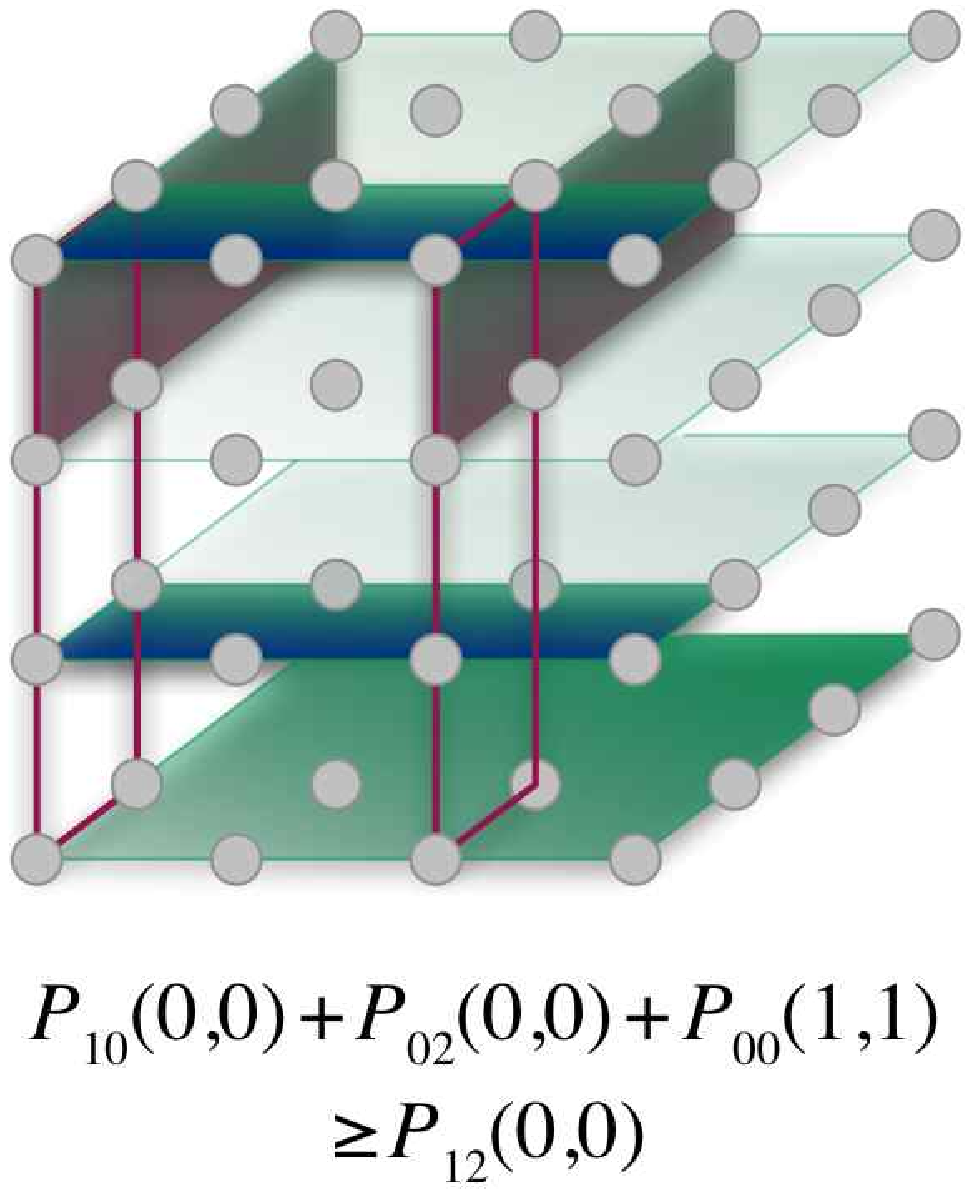}}
\label{fig6}
\caption
{
Graphical depiction of two-body three-axes Bell inequality which leads to ``Bell's'' Bell inequality.  Marginal probabilities $P_{ab}(A,B)$ are represented as identically colored planes for terms in LHS, a transparent square for the RHS, which are superimposed on the grid representing the underlying probabilities $V_{i,j,k}$ to be summed up.}
\end{figure}
\section{Discussion}
The derivation of Bell inequalities with the use of underlying probabilities examined here is quite general.  The argument is very straightforward, and this approach, in principle, is extendable to Bell inequalities with four or more choices and also to four or more players.  But actual graphical representation quickly becomes messy and intractable for higher number of choices and players, and therefore, is not expected to be competitive against traditional systematic approaches \cite{ZB02}.  There is, however, an advantage in our approach of having intuitive graphical representation, which should not be missed in pedagogical settings.

\bigskip
We thank Prof. I. Tsutsui and Dr. T. Ichikawa for stimulating discussions. This research was supported  by the Japan Ministry of Education, Culture, Sports, Science and Technology under the Grant number 15K05216.


%



\end{document}